# Policy-Carrying, Policy-Enforcing Digital Objects

Sandra Payette and Carl Lagoze

Department of Computer Science, Cornell University,
Ithaca, NY, USA
{payette,lagoze}@cs.cornell.edu

**Abstract.** We describe the motivation for moving policy enforcement for access control down to the digital object level. The reasons for this include handling of item-specific behaviors, adapting to evolution of digital objects, and permitting objects to move among repositories and portable devices. We then describe our experiments that integrate the Fedora architecture for digital objects and repositories and the PoET implementation of security automata to effect such object-centric policy enforcement.

## 1  From Traditional Access Control to Flexible Policy Enforcement

Access control in digital libraries must be scalable, flexible, and extensible – accommodating a wide range of objects and usage scenarios. This stands in contrast to traditional access control models, which are limited to a relatively fixed set of operating system abstractions (e.g., files) and computing actions (e.g., read, write). The limitation of these models has motivated the programming language, operating systems, and security communities to develop alternative and richer methods, including an approach known as language-based security [1].

The shortcomings of traditional access control schemes are particularly acute in digital libraries where idiosyncratic and complex digital objects are the norm rather than the exception. Such objects may aggregate mixed content (e.g., text, images, audio), may be highly interactive (e.g., a distance learning module), or may be portable (e.g., an object that is disengaged from its "home ship" for use on a mobile device). A typical security model has been to allow/disallow access to particular objects by particular subjects using system-level enforcement mechanisms. These mechanisms do not easily adapt to handle the wide range of scenarios pertinent to particular applications and complex objects. Designed to express only a limited set of access control rules, such mechanisms are often constrained in their *policy enforcement* capabilities.

In this work, we recognize a continuum of policy enforcement – ranging from repository or system-centric to object-centric. This model accommodates general system-wide requirements, such as those covered by standard access control mechanisms, along with more refined requirements that pertain to various types of content objects.

This spectrum can be understood through the metaphor of a museum attempting to enforce a variety of policies to protect its assets. In this hypothetical museum, a security guard is employed to stand guard at the main entrance and make decisions about who can enter for what reasons. The guard enforces some general rules of entry (e.g., people must have paid an entrance fee, people must check their bags in the cloak room), as well as policies that ensure the safety of the building (e.g., anyone carrying a weapon is denied entry). The guard may permit access to private rooms (e.g., by providing a special pass), but once a person enters the room, the guard cannot control what the person does. In terms of our continuum, the guard is a mechanism for enforcing museum-wide rules, typical of the system-centric end of our spectrum.

Carrying our metaphor forward, what happens once people have passed through the main entrance? We wouldn't expect the guard to have the knowledge or ability to oversee specific activities that occur inside the many rooms of the museum. Each room may have its own protocols, rules, and restrictions for specific behaviors. Also, room-specific rules may evolve on a day-to-day basis, or they may address the occurrence of special events. To be effective, the main guard would have to be omniscient and omnipresent, and equipped with all of the rules and tools for every unique situation.

As a better alternative, the guard should delegate control to others who enforce room-specific policies locally. A special exhibition manager would be equipped to enforce policies unique to a particular exhibition room. For example, she would monitor activity, allowing (1) members of the general public to touch certain items if they wear a white glove, (2) scholars with proper credentials to gain entrance to a private alcove to see selected rare documents, and (3) curators to open display cases to insert or remove items. Such policy enforcement is both object and context-specific.

Access control in digital libraries presents a challenge similar to that of the museum. Digital libraries contain a rich array of digital objects for which they must provide secure enforcement of both general-purpose and context-specific policies. This paper describes our work at Cornell to provide such secure policy enforcement.

## 2 Research Context

Our investigation of policy enforcement for digital libraries builds on and extends two ongoing research programs at Cornell: (1) distributed architectures for digital objects and repositories, and (2) enforceable security policies.

In the Fedora project, we have demonstrated interoperability and extensibility for digital objects and repositories [2,3]. The main result of this work is a flexible digital object model – one that supports the aggregation of mixed digital content and the association of rich and customized behaviors with those aggregations. For example, book objects have behavioral access points such as "Get Table of Contents" and "Get

Page." Multimedia learning objects have access points such as "Get Lesson Video." While Fedora enables the creation of such customized access points on objects, it still provides for interoperability among varying objects. Through a set of *generic* behaviors common to every digital object, Fedora enables clients to uniformly discover and invoke the customized, content-specific behaviors associated with the object.

Having demonstrated content-specific and extensible behaviors for digital objects, our next goal is to devise an equally flexible access control scheme to support them. The requirements for such a solution are:

- Flexible policy definition: ability to define policies that are either general-purpose or specific to particular digital object behaviors
- Highly secure enforcement: policy enforcement mechanism must be based on well-grounded security techniques
- Conducive to mobile code and mobile computing environment: policies must be enforced on any code that is obtained from the network. Furthermore, policies and enforcement mechanisms must permit digital objects to move among repositories and to mobile devices.

This paper describes an approach for fulfilling these requirements by applying and extending new work in language-based security. Specifically, we are investigating the use of *security automata* for specifying access control policies that are flexible, fine-grained, and *enforceable* [4]. This special class of automata can be used to model policies in the form of finite-state machines. For example, using a state transition diagram, Figure 1 depicts an automaton for the policy "after viewing descriptive metadata, only Cornellians can access the Lesson 1 video (of a distance learning module)."

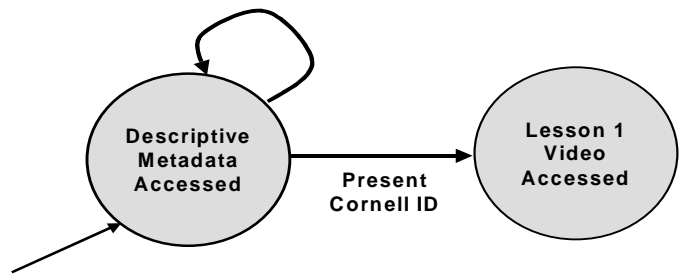

**Fig. 1.** Simple Security Automata for a Lecture Object Policy

Security policies modeled in this manner can be enforced by mechanisms that simulate an automaton. Essentially the automaton will monitor the executions that occur in a target system (including application-level and object-level executions). It will prevent the running of any executions that violate a system policy. This approach is demonstrated in Cornell's Policy Enforcement Toolkit (PoET). In PoET, security automata simulations are implemented as in-line reference monitors (IRM) for the Java environment [5,6]. Through our experiments with PoET and Fedora, we have demonstrated the power of IRM in providing flexible and secure access control for behaviorally-rich digital objects and repositories. The details of the Fedora and PoET technologies will be described in Section 4.

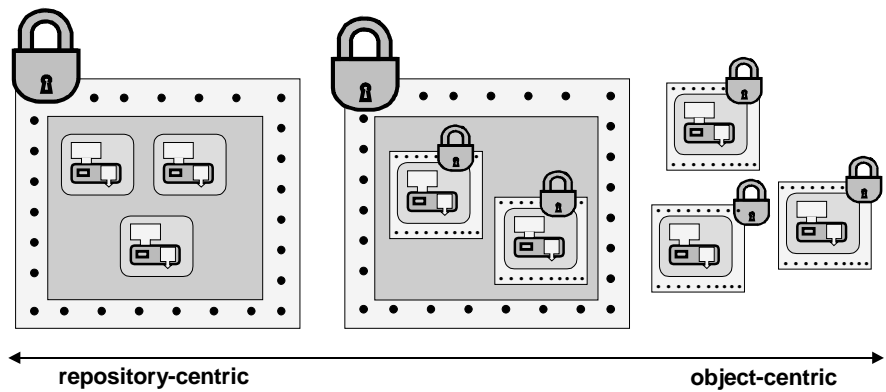

**Fig. 2.** Policy Enforcement Continuum

As illustrated in Figure 2, our integration of the Fedora and PoET technologies can provide policy enforcement for a range of digital library scenarios. We depict a policy enforcement continuum (from repository-centric to object-centric) where the dotted areas with "locks" indicate the access control target points. On the left, a Fedora repository enforces general-purpose policies over all its contained objects. In the middle, a hybrid approach is used, where some policy enforcement occurs at the repository, while some policies apply to individual digital objects. At the right end of the continuum are digital objects that handle their own policy enforcement. The experiments we report on in this paper are those focused toward the object-centric end of the spectrum. To demonstrate this object-centric approach to policy enforcement, we have designed the *Policy-Carrying, Policy-Enforcing Digital Object*. The next section discusses the motivations for pushing policies and enforcement mechanisms into digital objects.

# 3 Policy-Carrying, Policy-Enforcing (PCPE) Digital Objects

Most digital libraries will employ a combination of repository-level and object-level policies. The following requirements indicate that the object-level end of the spectrum deserves special attention.

## 3.1 Behavior-Oriented Policies

Policy definition should be able to address the characteristics, behaviors, and usage scenarios of actual digital library objects - not just system abstractions such as files, directories, and operating system commands. Our Fedora work demonstrates a model whereby behaviors of digital objects correspond to behaviors "real-world" entities. These behaviors are implemented as methods that essentially hide the complexities of underlying data, structure, and file formats that are encapsulated in a digital object container. Referring to our book example, methods such as "Get Table of Contents", and "Get Page" provide access to content, whether a book is stored as a postscript file, or as a set of digital images.

In Fedora, clients invoke such methods to obtain different "views" of digital objects. To control access to digital object content, we must be able to write policies that correspond to these defined views. In other words, the semantics of policies should parallel the semantics of digital object behaviors. This allows policy definition to be closely aligned with people's understanding of their content, not of a computer system.

## 3.2 Specificity vs. Generalization

Although we may be able to identify certain repeatable "design patterns" for policies in digital libraries, we should recognize that particular combinations of these types of policies are often unique to particular objects. This corresponds to the fact that digital resources are generic in some ways, but specialized in other ways.

There are circumstances in which policies should be applied to digital objects in a generic manner. Consider a repository that has a variety of digitized books and lectures pertaining to computer science. There may be a set of *general-purpose* policies pertaining to the collection as a whole. An example of such a policy is: "Cornellians with repository manager status can add and delete objects in the collection." This policy is quite coarse in nature, and applies equally to all objects in the repository regardless of their individual characteristics. These policies are typically enforced by a repository-wide mechanism.

However, repository-level policy enforcement becomes unworkable for policies that address the idiosyncratic nature of individual digital objects. General-purpose enforcement mechanisms may not be expressive enough to support the variety of vari-ables, conditions, and desired actions pertaining to particular content items. As an example of such an item-specific policy, consider an object that provides access to a digitized lecture. Table 1 outlines a plausible policy governing access to such an object.

**Table 1:** Policy for Lecture Object "A"

**Policy Rules:**
- Only users who present Cornell credentials, or who pay a $5 fee, can access the HIGH-RESOLUTION VIDEO.
- All Users can access SLIDES 1-10 without restriction.
- Only users who present Cornell credentials can access SLIDES 11-20.
- All users can access DESCRIPTIVE METADATA without restriction.

Enforcement of such a policy requires a mechanism that can detect requests that correspond to the access points specific to the lecture object, and ensure that the conditions of the policy are not violated. Although one could attempt to extend repository-level mechanisms to deal with these kinds of policies, it is probable that the mechanism would eventually becomes awkward and overburdened as it tried to accommodate increasing numbers of customized policies. Such object-specific policies are better dealt with in a modular fashion, and linked to the object and its individual access points, instead of a repository.

### 3.3 Decentralized Policy Management

We envision digital libraries with distributed responsibility for policy definition. In this scenario, system managers implement and maintain policies to ensure security for entire repositories. However, other information stewards have responsibility for defining and maintaining usage policies for sets of digital objects. As we move policy definition to the object level, we can delegate responsibility for policy management to agents who are very familiar with certain classes of objects, or who have a vested interest in managing certain collections.

A critical component of such distributed policy definition is a high-level policy language. There are a number of ongoing projects that are developing such languages, including Cornell's PoET, the XrML initiative [7], and PolicyMaker [8]. From a digital library perspective, such languages must be intuitive and easy, so that authors and librarians can write policies tailored to particular digital objects, or groups of objects.

### 3.4 Extensibility

If we take extensibility as a key requirement for digital objects, we are presented with the challenge of evolving policies with the evolution of an object. One of the advantages of the Fedora architecture is the ability to enhance existing objects, over time, by endowing them with new behaviors. Such behavior evolution is facilitated by the modular design of Fedora Digital Objects, and particularly by the use of "pluggable" mechanisms that can add or transform object behaviors. The principle of modularity should be applied to policies and enforcement mechanisms as well. When object behaviors or interfaces change, policies and enforcement mechanisms must easily adapt. Attempts to define and manage monolithic policies at the repository-wide level would likely be difficult, especially since we cannot anticipate ahead of time the manner in which policies may need to be changed to support new object-level behaviors.

### 3.5 Portability and Mobile Computing

Portability of digital objects is important for a number of reasons. First, fast and reliable access to objects at a global level often requires caching or mirroring. Second, as we move from a world of fixed networks to one of mobile devices with intermittent connectivity (nomadic devices), we will increasingly confront issues of access to digital content that is copied out of the networked infrastructure. In both cases, we would like digital objects to be secure in transit and to facilitate their own policy enforcement at their target destinations.

## 4   Experiments: PCPE Digital Objects using Fedora and PoET

Given the motivations just described, we devised set of experiments to demonstrate Policy Carrying, Policy-Enforcing (PCPE) Digital Objects. In preparation for the experiments, we evolved the Fedora model to support an augmented Digital Object - one that carries its own policies within. As mentioned earlier, the security automata approach provides significant flexibility for policy definition and enforcement, making it an appropriate technology for addressing the unique requirements of digital libraries. Thus, by extending Fedora, and applying Cornell's PoET software, we have stretched the capabilities of our digital object architecture, and have simultaneously tested the flexibility of PoET against real-world requirements of the digital library community. A brief review of the both Fedora and PoET will help illuminate the PCPE experiments.

### 4.1   Technical Overview: Fedora

Fedora is a modular architecture built on the principle that interoperability and extensibility is best achieved by the clean separation of data, interfaces, and mechanisms [2,3]. A Fedora Repository provides a general-purpose management layer for *Digital Objects*. In their simplest form, Digital Objects are containers that aggregate mime-typed streams of data (e.g., digital images, XML files, metadata), known as *DataStreams*. It should be noted that DataStreams can be references to external data - either disseminations of other Fedora Digital Objects, or service requests to remote data sources. These are known as *ReferenceDataStreams*, and are one way that Fedora enables the aggregation of distributed content. Clients can interact with Fedora Digital Objects through a generic set of methods, known as the *Primitive Disseminator*. These methods can be thought of as the default API for Digital Objects, providing a well-defined open interface for all objects.

In addition to behaving in a generic manner, Digital Objects must be able to mirror real-world entities by providing access methods that make an object behave in a content-specific manner. For example, a natural behavior for a book would be "Get Table of Contents." Fedora allows the association of rich and extensible behaviors with Digital Objects by "plugging in" generic components known as *Disseminators*. Each Disseminator will aggregate references to: (1) a formally defined behavior interface that defines a set of methods for a particular kind of digital library resource (e.g. Book interface, Lecture Browser interface), (2) an executable mechanism (i.e., a Servlet) that runs these methods, and (3) DataStreams the Servlet should use to fulfill specific method requests. These interfaces and mechanisms are themselves disseminated by Digital Objects, laying the foundation for unlimited extensibility of the architecture. A major strength of the Fedora extensibility model is that clients can use the generic methods (of the default API) to discover and invoke content-specific methods defined on Disseminators. The Digital Object facilitates the invocation of these extended methods, returning customized disseminations of content to the client.

Figure 3 depicts a Digital Object that provides access to a multimedia lecture. There are multiple DataStreams in the object's core: a low-resolution video of a lecture. (Video-L); a high-resolution video (Video-H); a set of accompanying slides (Slide-1, Slide-2); and an XML file that encodes both synchronization and descriptive metadata. Two Disseminators define the access methods for this object. One defines an interface for experiencing the lecture, the other an interface for obtaining Dublin Core metadata. Although it is not apparent in the diagram, each of these Disseminators references a Servlet - a piece of java code that may be local to the repository, or may be a piece of mobile code. Clients interact with the Digital Object through its generic API methods.

Providing access control for these complex objects, without compromising flexibility and extensibility, presents a significant challenge. Early Fedora designs specified Digital Objects with their own internal access control mechanisms, known as Access Managers. The Access Manager was envisioned as a watchdog, overseeing the execution of all methods that could be invoked on a particular Digital Object. The power and flexibility of the PoET software has made it possible for us implement the Access Manager abstraction using the sound underpinnings of security automata theory. An overview of the PoET technology lays the groundwork for the rest of this section.

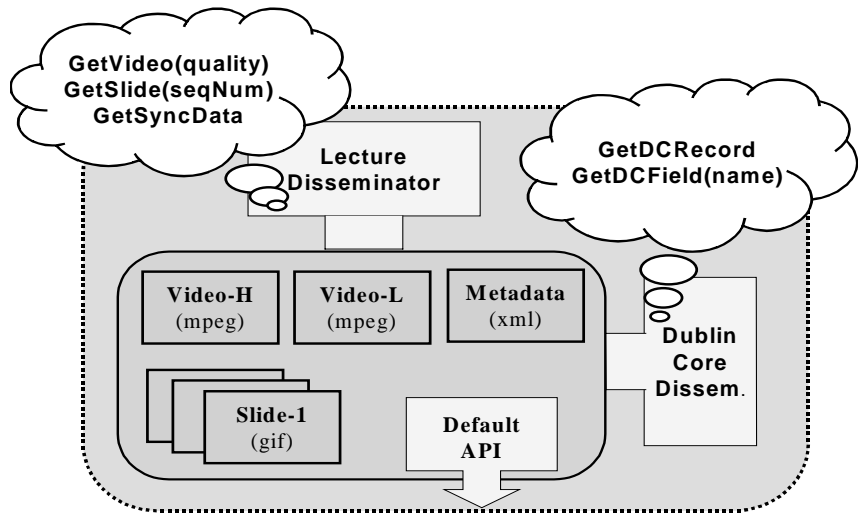

**Fig. 3.** A multimedia lecture represented as a Fedora Digital Object

### 4.2 Technical Overview: Policy Enforcement Toolkit (PoET)

PoET is an implementation of in-lined reference monitors (IRMs) for Java applications. To review, IRMs can simulate security automata; as such, they can enforce a rich class of access control policies. In general, a reference monitor enforces security policies by mediating all executions that pertain to the policy it is enforcing. A reference monitor must be transparent to applications, and must be protected from attack by applications. IRMs achieve this through bytecode modification by a trusted program rewriter.

In PoET, java applications are converted to *secured* applications by a code rewriter that embeds checks into Java Machine Language programs (bytecode). An easy way to understand this is to imagine PoET "baking" the bytecode in such a manner that the policy becomes ingrained within it. The embedded checks terminate forbidden executions *before* they violate the security policy for the application. To ensure the integrity of the checks, and to increase performance of the modified code, PoET performs program analysis and optimization during the rewriting process. It should be noted that since PoET modifies bytecode - *not* Java source code - the Java compiler is *not* part of the trusted computing base. It follows, then, that a computer running PoET does not require the presence of a Java compiler, nor Java source code for the application.

In PoET, security policies are specified in the *Policy Specification Language* (PSLang), which is a simple, event-oriented language that is somewhat similar to Java in syntax. Policies that are modeled as security automata can be easily expressed in PSLang.

Figure 4 shows the process by which PoET creates a secure application. First, PoET intakes both the policy to be enforced, and the code to be protected. The PoET rewriter decomposes bytecode (Java class files), embeds policy-checking code, performs integrity checks, and generates a modified versions of the class files. All original functionality is preserved, but the resultant class files have the policy checks merged into them. Finally, the PoET class loader transfers the modified code into the Java Virtual Machine (JVM) where it is run as a secured application.

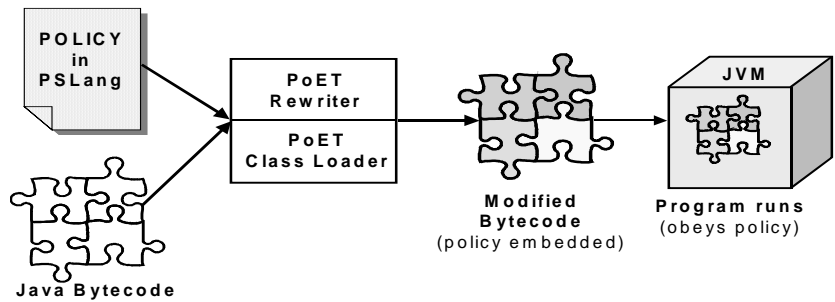

**Fig. 4.** PoET embeds a policy into Java bytecode

### 4.3 Integration of Fedora and PoET

Because the Fedora implementation is written as a Java application we were able to take full advantage PoET's in-line reference monitoring. The standard model for PoET is to statically embed policies into application code. Since Digital Object behaviors may evolve, the policies that correspond to them must also evolve. If policies are modified or extended, these changes must be immediately reflected in the executions of Fedora Digital Objects. To ensure this, we enhanced PoET so that it now accepts and modifies pieces of Java bytecode on a "just-in-time" basis. This was particularly important because Digital Objects often depend on shared pieces of mobile code to execute behaviors.

Fedora now associates a *default policy* with each Digital Object in a Repository. The default policy is a *shared* policy that any Digital Object can point to. The default policy specifies access restrictions for the generic API methods common to all Digital Objects. Typically, repository managers will define a default policy that prevents end-users from modifying the contents of Digital Objects, but allows users to browse and obtain disseminations from them.

Additionally, Fedora now supports the association of one content-specific policy with each behavior interface (e.g., a book interface that includes a number of methods).

These policies can be customized to a particular Digital Object, or they can be "group policies" that any Digital Objects can associate with. Customized policies are stored as DataStreams inside Digital Objects. Group policies reside in the repository and are stored *by reference* in Digital Objects (i.e., stored as ReferenceDataStreams).

To provide policy enforcement using in-line reference monitoring, Fedora interfaces with PoET. In Fedora, extensible code modules (i.e., Servlets) are used to implement Digital Object behaviors. Originally, Fedora's *Mechanism Manager* obtained these code modules – either locally or from the network - and would load them into the JVM as needed. As part of our integration, we adapted the Mechanism Manager to interface directly with PoET. Now, whenever a Digital Object receives a request, the Mechanism Manager obtains the requisite code module and sends the appropriate policy and code to the PoET rewriter. From there, PoET takes over to ensure the secure modification of the Java bytecode, and secure loading of the modified classes into the JVM.

### 4.4 Experiment Results

In collaboration with our library colleagues, we are designing policy enforcement experiments that address policy requirements for a range of object types and institutional contexts. In this paper, we report on our first set of experiments with Cornell's Computer Science lecture archive. Figure 5 depicts a lecture packaged as a Fedora Digital Object secured with PoET's in-line reference monitoring. It should be noted that the diagram does not convey the fact that repository-wide policies are normally in place to prevent untrusted programs from compromising the host computing environment. In the diagram, the Lecture Digital Object introduced earlier now contains two policies: (1) a policy pertaining to the Digital Object's default API behaviors (labeled "Default Policy"), and (2) a customized policy pertaining to the behaviors of the Lecture Disseminator (labeled "Policy-L"). The default policy restricts access to operations that can modify the contents of the object. The Lecture Disseminator policy (Policy-L) is a PSLang specification of the policy we saw earlier in Table 1. Although other lecture objects are structurally similar, they contain different content and different policies.

In support of the object-centric view we have proposed herein, we have achieved three main results, namely the ability to: (1) write highly customized policies tailored to the specific nature of object content, (2) enforce radically different policies for different objects of the same type, and (3) dynamically apply policies to code modules (mobile java programs) that are shared by multiple objects, demonstrating that the same application code can adapt to the policies pertaining to a particular run-time context.

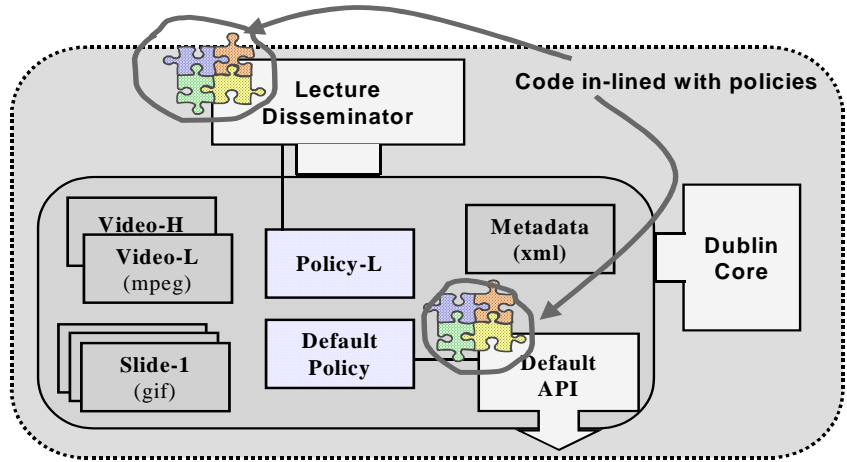

**Fig. 5.** "Lecture A" Digital Object with Restrictive Policies

## 5 Discussion

Object-centric policy management may have its drawbacks. Is it too hard to associate policies with every object? It depends on the nature of the objects, how unique they are, and how uniform policies tend to be. However, as noted earlier, repository-centric policy management has serious scalability and extensibility issues. Our future experiments will focus on the trade-offs of each.

In addition, there are significant mobile computing challenges yet to be addressed with PCPE digital objects. What if a digital object is taken out of its home repository and downloaded onto a mobile computing device? Such objects have moved outside our trusted computing base. For such *portable* objects there is a level of assumption around the "receiving" environment. First, the receiving environment must be able to interpret and "run" the digital object. In our experiments, we assume the presence of a Java Virtual Machine and software capable of interpreting Fedora objects. Even if policies are embedded within bytecode, an object may still be vulnerable. A portable digital object must be tamper-proof. In our Fedora/PoET implementation, a digital object that is disembodied from its home repository could be compromised by a savvy attacker who knows how to decompose and modify java bytecode in a manner that allows a policy to be bypassed. Although this is a difficult maneuver, we believe that our PCPE digital objects should be supported by supplemental security measures - possibly the use of public key infrastructure or other trust schemes to support digital object mobility. The digital objects should be secured containers, similar to that approach taken by IBM Cryptolopes [9]. Trust and security for mobile computing are active research areas that will be a continuing thread of inquiry in our digital library research.

### 5.1 Future Work

Under the auspices of Project Prism [10], we will continue to test increasingly complex policy scenarios in collaboration with Cornell University Library. One area of focus will be in contention between repository policies and digital object policies - how does the overlay or interaction of policies get resolved? A particularly interesting line of future research will be in the area of "intentional" policies. At this time, we have demonstrated the pre-defined binding of particular policies to particular digital objects. In our future work we will study dynamic binding of policies to "virtual subsets" of digital objects. We look to dynamic rule-based policy binding where criteria can be established that set up the conditions in which objects should be subject to particular polices. This would allow us to say at some arbitrary point in time that "all objects authored by Bill Arms are now subject to the provisions of Policy XYZ."

### 5.2 Related Work

The Smart Object Dumb Archive (SODA) architecture promotes a similar notion to our PCPE Digital Objects by pushing functionality into objects [11]. Although SODA objects may contain terms and conditions data, it is not clear whether these are expressed in a formal policy language, or what security techniques are used to enforce them. Also, the SODA work does not specifically address issues of flexibility in policy expression and adaptability of enforcement mechanisms.

Others have developed policy specification languages that can be used in conjunction with trust management systems. Keynote [12] has its own policy language that is used in conjunction with the Keynote interpreter. XrML [7] is an XML-based markup language for describing rights, fees, and terms for digital content. XrML does not prescribe an enforcement mechanism, but a way to specify rights information so that various systems can read and interpret it. Cornell's PoET is an application of language-based security. There is no central trust server that grants or denies access. PoET's policy language (PSLang) is optimized for use by a program rewriter that actually modifies application code (java bytecode) to ensure that policies are enforced. The modified programs will prevent executions that violate policy. Relative to other approaches, this technique is particularly useful in securing mobile code applications.

### Acknowledgements:

We acknowledge the contributions of Fred Schneider for security automata theory and PoET software; Ulfar Erlingsson for PoET software; Naomi Dushay for work on the

Fedora reference implementation; Sugata Mukhopadhyay and Jill Newman for the lecture object testbed. Work described here was supported by NSF Grant IIS-9817416 and ARPA/RADC grant F30602-96-1-0317, AFOSR Grant F49620-94-1-0198, DARPA and Air Force Research Laboratory, Air Force Material Command, USAF, under agreement number F30602-99-1-0533, NSF Grant 9703470, and a grant from Intel Corporation. The views and conclusions herein are those of the authors and should not be interpreted as necessarily representing the official policies or endorsements, either expressed or implied, of these organizations or the US government.